\documentclass[natbib,twocolumn]{article}
\usepackage[a4paper, total={7in, 10in}]{geometry}
\pdfoutput=1

\usepackage{siunitx}
\usepackage{mathrsfs}
\usepackage{amsmath}
\usepackage{amssymb}
\usepackage{amsfonts}
\usepackage{gensymb}
\usepackage{dblfloatfix} 

\renewcommand{\mathbb}[1]{\mathbf{#1}}
\everymath{\displaystyle}

\usepackage{interval}
\usepackage[pdftex]{graphicx}
\usepackage[svgnames]{xcolor}
\graphicspath{{figures/}}

\usepackage{multirow}
\usepackage{booktabs,tabularx}
\usepackage{makecell}
\usepackage{array}

\title{Storm Surges as Seen by Coastal and Spaceborne Radars:\\Case Studies in British Columbia}

\author{Baptiste Domps and Charles-Antoine Gu\'erin
	\thanks{B. Domps is with Degreane Horizon, 83390 Cuers, France (e-mail: baptiste.domps@degreane-horizon.fr). C.-A. Gu\'erin is with the Mediterranean Institute of Oceanography (MIO), Universit\'e de Toulon, Aix-Marseille Universit\'e, CNRS, IRD, Toulon, France (e-mail: guerin@univ-tln.fr).}
}

\begin{document}

\maketitle

\begin{abstract}
  Short-term sea level fluctuations prompted by abrupt atmospheric changes can be hazardous phenomena for coastal regions. We report on two such recent storm surges that occurred in 2020 on the shores of British Columbia, Canada. A rare concordance of ground-based and spaceborne sensors made it possible to observe these events with a variety of instruments : (1)~a coastal oceanographic radar; (2)~the synthetic aperture radar onboard satellite Sentinel-1B; and (3)~a network of shoreside tide gauges. In the light of these case studies we show how satellite-based radar data can be used to complement the observation and interpretation of ground-based measurements in the context of ``tsunami-like'' sea level oscillations.
\end{abstract}

\section{Introduction}

\begin{figure}[h]
	\centering
	\includegraphics[width=.7\linewidth]{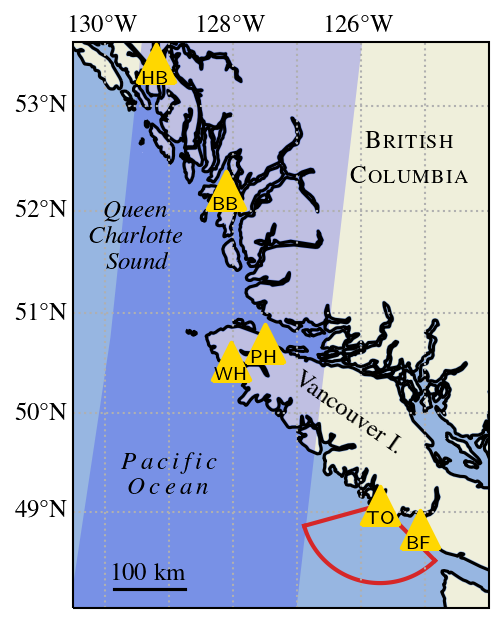}
	\caption{Map of the study area along the west coast of BC. Selected CHS tide gauge stations are represented as {\color{Gold}\large$\blacktriangle$} (HB: Hartley Bay; BB: Bella Bella; PH: Port Hardy; WH: Winter Harbour; TO: Tofino; BF: Bamfield). Superimposed is the footprint of Sentinel-1B observations used in this study. Typical coverage of the HFR of Tofino is also shown as {\color{red}\bfseries ---}.}
	\label{fig:map}
\end{figure}

Coastal High-Frequency Radars (HFR) are today widely used for the real-time monitoring of sea surface currents (e.g. \cite{roarty2019}). Ocean Networks Canada (ONC) has been operating one such instrument since 2015 in Tofino, on the West Coast of Vancouver Island, British Columbia (BC), Canada (Fig. \ref{fig:map}). This HFR is equipped with a tsunami warning software based on the detection of abnormal surface currents patterns \cite{dzvonkovskaya2018}. The first real-time warning, on October 14, 2016, occurred in the absence of any reported seismic activity and was therefore related to the family of atmospheric-driven phenomena (\cite{dzvonkovskaya2017,guerin2018}). Unlike seismic tsunamis, which are generated by earthquakes or landslides, such events are induced by strong atmospheric disturbances (e.g. pressure jumps, atmospheric fronts, etc. \cite{monserrat2006}). The atmospheric forcing can trigger two kinds of ``tsunami-like'' sea-level oscillations: long ocean waves (known as ``meteorological tsunamis'') or storm surges. The former yet requires some form of resonance between the atmospheric propagation speed and the oceanic long wave phase speed. Both can produce damages along the seashore, demonstrating the need for early warning systems in the risk zones \cite{vilibic2016} such as the shore of British Columbia \cite{thomson2009}. In this paper we will only consider the detection of storm surges, which are more common than meteo-tsunamis.

The main point is that the coastal HFR systems that are used for tsunami warning can also be utilized for the detection and warning of storm surges. While the first observation of a storm surge using HFR occurred in the eighties \cite{lipa1986}, few observations were reported until the past few years (e.g. \cite{heron2022,dompsPhD}). In particular, the October 2016 event was recently reanalyzed by the authors, who shown that it was more likely a storm surge than a meteorological tsunami \cite{domps2022}. It now seems clear that conventional tsunami detection algorithms can be triggered by abrupt changes in surface currents propagating towards the coast, should they be of seismic or atmospheric origin. Such ability makes HFR irreplaceable for the short-term storm surge coastal warning. Thus, it was recently proposed by \cite{heron2022} to use the wind direction measured by HFR to discriminate between tsunami waves and storm surges in real-time. However, this emerging application still requires cautious assessment in the light of synthetic and real test cases. In this paper, we analyze two tsunami-like events that caused a warning to be issued by the HFR of Tofino in 2020, on January 5 and November 23. Both were fortuitously observed by the synthetic aperture radar (SAR) onboard satellite Sentinel-1B during its flyby of BC, just before the warnings (Section \ref{sec:sar}). To our best knowledge, this would be the first reported observation of a storm surge in open ocean using spaceborne SAR. We compare the event arrival times derived by the HFR of Tofino and a network of coastal tide gauges (Section \ref{sec:ground}) and show that, similarly to the October 2016 event, both warnings were more likely caused by a storm surge rather than a meteorological tsunami (Section \ref{sec:discussion}). 

\section{Tide Gauges and HF Radar Measurements}\label{sec:ground}

\begin{figure}[p]
	\centering
	{
	\includegraphics[width=\linewidth]{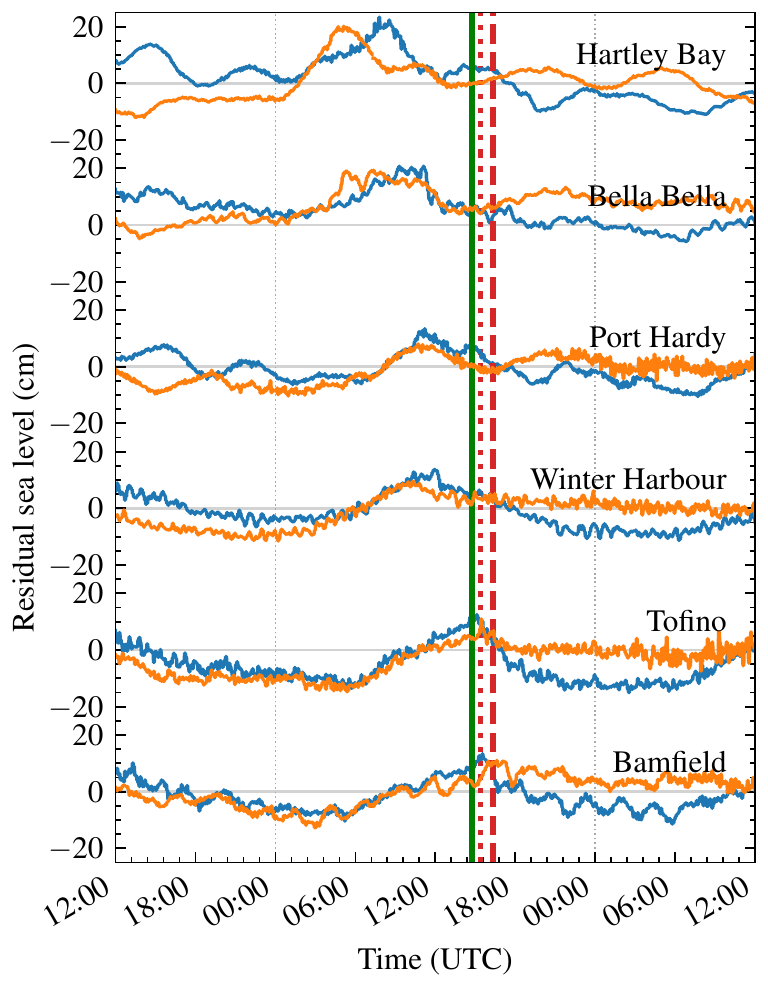}
	\caption{Sea level anomalies at 6 CHS tide gauges (\si{\cm}; see Fig.~\ref{fig:map}) from: ({\color{SteelBlue}\bfseries ---})~Jan. 4 to 6, 2020; and ({\color{DarkOrange}\bfseries ---})~Nov. 23 to 25, 2020. The times of the Sentinel-1B flybys (at 14:43~UTC on Jan. 5 and 14:45~UTC on Nov. 24, {\color{Green}\bfseries ---}\,; see Fig. \ref{fig:sar}) and of the tsunami alerts issued by the HFR of Tofino (at {\color{FireBrick}\bfseries $\cdot\cdot\cdot\cdot$} 15:22~UTC on Jan. 5 and {\color{FireBrick}\bfseries -\,-\,-} 16:20~UTC on Nov. 24) are also shown.}
	\label{fig:tides}
	}
	\vspace{5mm} %
	{
	\includegraphics[width=\linewidth]{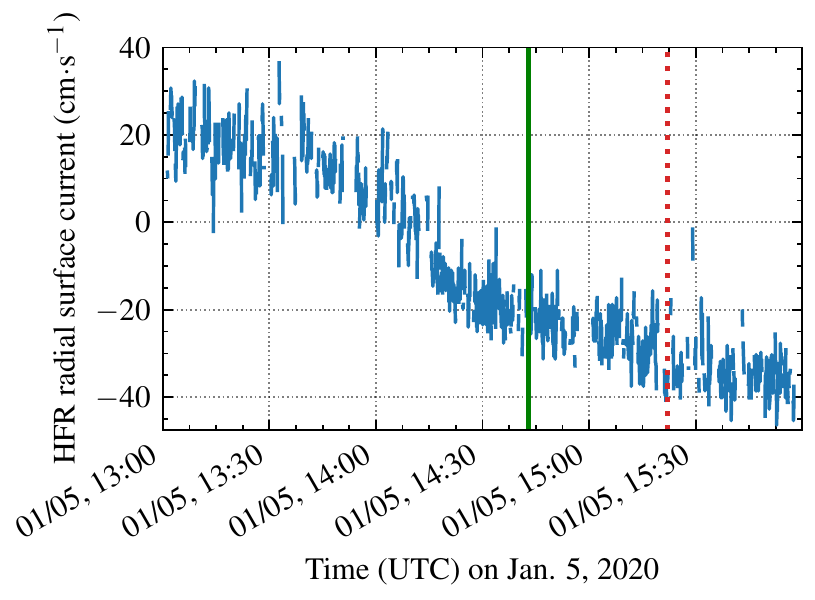}
	\caption{Time series of radial surface current (\si{\cm\per\s}) inverted in the middle of the beam of the HFR of Tofino (Fig. \ref{fig:map}) for the Jan. 2020 event and computed using the TVAR-MEM approach \cite{domps2021a,domps2022} every \SI{4}{\s} from overlapping signals of \SI{33}{\s}. Negatives values represent surface currents directed toward the radar.}
	\label{fig:ur}
	}
\end{figure}

The shores of BC are equipped with tide gauges from the Permanent Water Level Network (PWLN), operated by the Canadian Hydrographic Service (CHS) \cite{rabinovich2004}. Predicted tides were removed from the measurements and the resulting sea level anomalies within the study area are shown in Figs. \ref{fig:tides}a and \ref{fig:tides}b for the Jan. and Nov. 2020 events, respectively. In both cases the records are very similar and show sudden rises in sea level, up to \SI{20}{\centi\m}. The time shift between the sea level maxima suggests that both phenomena were induced by an atmospheric disturbance propagating from the northwest to the southeast at an average speed of 15 to \SI{20}{\m\per\s}, possibly being a maritime polar (mP) air mass originating from the Gulf of Alaska.

In coastal regions the main amplification mechanism of atmospheric waves at the sea surface is the so-called Proudman resonance \cite{monserrat2006}, occurring when the atmospheric propagation speed equals the ocean waves speed $c=\sqrt{gh}$, where $h$ is the local depth and $g$ is the gravitational acceleration. Thus, the observed atmospheric speeds would correspond to resonant depths of 25 to \SI{40}{\m}, which seems to be way too confined to set off a meteorological tsunami, considering the bathymetry off of BC \cite[Fig.~1]{domps2022}. This supports the hypothesis that the observed events were rather related to a storm surge. Propagation of the gusty polar air mass over the ocean induced a sudden shift in sea surface currents velocity and direction (of the order of a few tenth of \si{\centi\m\per\s}, Fig.~\ref{fig:ur}), triggering the HFR automatic tsunami warning in Tofino (at 15:22 and 16:20~UTC in Jan. and Nov., resp.). This is consistent with the warnings being issued just after the sea level maxima, as already noticed by \cite{heron2022} for the January event.

\begin{figure*}[!b]
	\centering
	\begin{minipage}[c]{.25\linewidth}
		\centering
		\includegraphics[width=\linewidth]{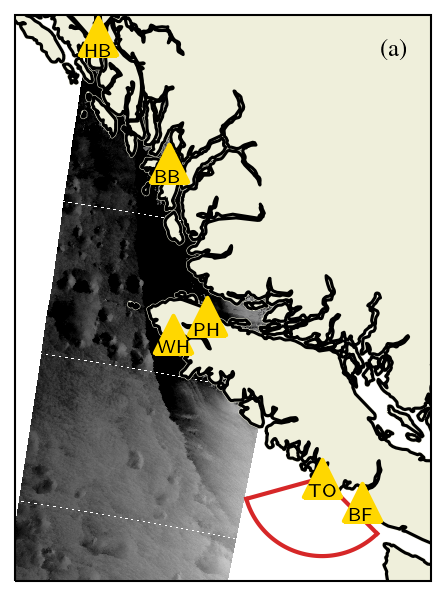}\\
		\includegraphics[width=\linewidth]{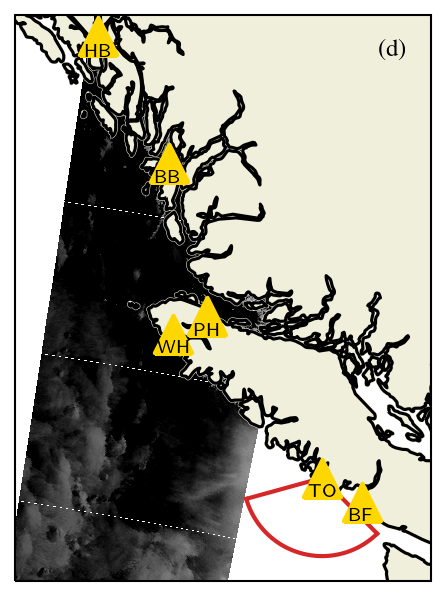}
	\end{minipage}
	\begin{minipage}[c]{.0825\linewidth}
		\centering
		\includegraphics[width=\linewidth]{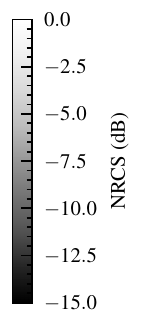}
	\end{minipage}
	\hfill %
	\begin{minipage}[c]{.25\linewidth}
		\centering
		\includegraphics[width=\linewidth]{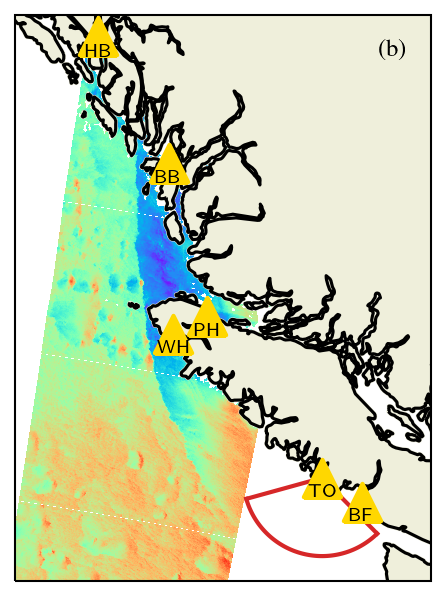}\\
		\includegraphics[width=\linewidth]{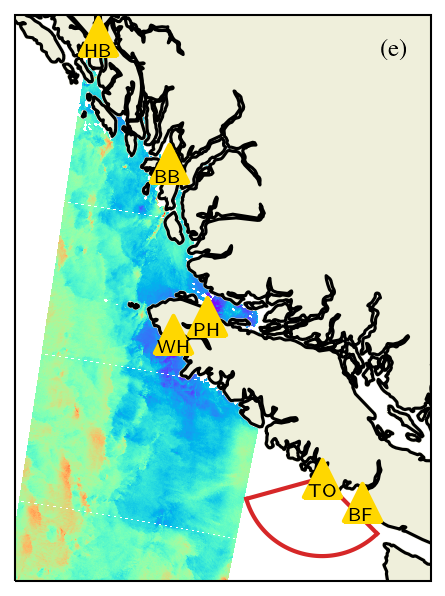}
	\end{minipage}
	\begin{minipage}[c]{.065\linewidth}
		\centering
		\includegraphics[width=\linewidth]{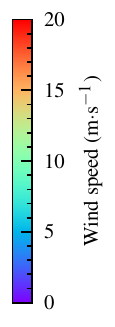}
	\end{minipage}
	\hfill %
	\begin{minipage}[c]{.25\linewidth}
		\centering
		\includegraphics[width=\linewidth]{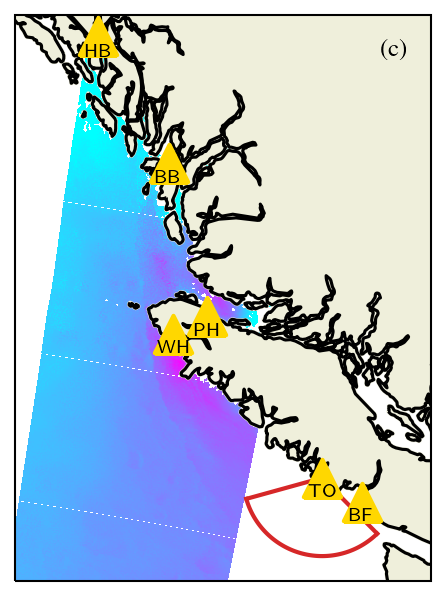}\\
		\includegraphics[width=\linewidth]{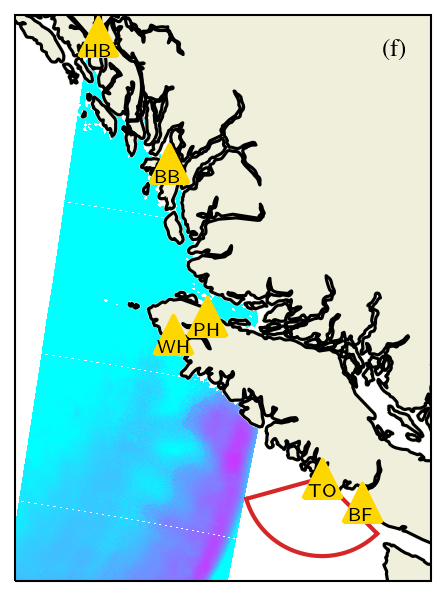}
	\end{minipage}
	\begin{minipage}[c]{.07\linewidth}
		\centering
		\includegraphics[width=\linewidth]{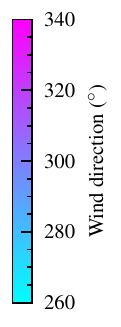}
	\end{minipage}
	\caption{SAR data retrieved from satellite Sentinel-1B flying over British Columbia (see Fig. \ref{fig:map}) in interferometric wide (IW) swath mode at: (top)~14:43~UTC on Jan. 5, 2020; (bottom)~14:45~UTC on Nov. 24, 2020. Left: VV-NRCS map (\si{\deci\bel}); middle: ocean surface wind speed $u_{10}$ (\si{\m\per\s}); right: ocean surface wind direction $\theta_{10}$ (meteorological convention, \si{\degree}).}
	\label{fig:sar}
\end{figure*}

\section{Sea Surface as Seen by Sentinel-1B SAR}\label{sec:sar}

Sentinel-1 is a constellation of two SAR satellites operating in C-band, as part of the European Copernicus earth observation program. Over the ocean, they provide all-weather, day-and-night images of the surface normalized radar cross section (NRCS) at the meter-scale resolution. In C-band, this NRCS corresponds to the sea surface roughness being mostly dominated by the local wind. Thus, numerous air-sea interaction processes can be observed in these images (e.g., \cite{wang2019}). However, while the first observation of a storm surge in open ocean using satellite altimeters goes back to 2005 \cite{scharroo2005}, there has been no such detection with SAR images. Fortuitously, the Jan. and Nov. 2020 events were observed by satellite Sentinel-1B just before the HFR warnings. The flybys occurred at 14:43 and 14:45~UTC, respectively.

The atmospheric front can be observed in the sea surface roughness images, where a high electromagnetic return indicates a rough surface associated to gusty winds. This is particularly visible for the Jan. 2020 event (Fig. \ref{fig:sar}a) where the two air masses are clearly distinct. We used the optimal inversion method \cite{portabella2002} with the CMOD7 geophysical model function (GMF) from the KNMI \cite{stoffelen2017} to invert the \SI{10}{\m}-wind speed ($u_{10}$) and direction ($\theta_{10}$) from the SAR images. The resulting maps (Figs. \ref{fig:sar}b, c, e and f) show the atmospheric process at the origin of the oceanic event. A sudden rise in $u_{10}$ (from 5 to \SI{15}{\m\per\s}) and shift in $\theta_{10}$ (from 270 to \SI{320}{\degree}, that is from west to north-west) is seen at the passing of both fronts. Considering that the average velocity of the wind-driven surface current is of the order of \SI{2}{\%} of $u_{10}$ \cite{berta2018}, this is consistent with the amplitude of the jump in radial surface current observed by HFR.

\section{Discussion and Conclusion}\label{sec:discussion}

Coincident data from Sentinel-1B, the HFR of Tofino and tide gauges have been synthesized to observe two storm surge events in January and November 2020. At the times of the flybys, the atmospheric fronts were located in the open sea according to the SAR images while the storm surges had already hit the shores of BC, according to the tides gauges and the HFR warnings. This indicates that the wind-induced bulges in sea level were located ahead of the atmospheric fronts. We also investigated the speed of the atmospheric fronts between the Sentinel-1B flybys and the HFR warnings. The distances between the fronts as seen in the SAR images and the western outer extent of the HFR coverage are of the order of \SI{50}{\kilo\m} (Jan. event) and \SI{100}{\kilo\m} (Nov. event) while the delay between the Sentinel-1B flybys and the HFR warnings are of \SI{39}{\minute} and \SI{1}{\hour}~\SI{35}{\minute}. This leads to an average speed of the order of \SI{75}{\kilo\m\per\hour} (\SI{21}{\m\per\s}) and \SI{65}{\kilo\m\per\hour} (\SI{18}{\m\per\s}), respectively, which is consistent with the typical speed of cold fronts. Further case studies will definitely be needed to assess a systematic detection of storm surges from spaceborne SAR images and HFR tsunami-warnings.

\section*{Acknowledgment}
\addcontentsline{toc}{section}{Acknowledgment}
First author was supported by the Direction G\'en\'erale de l'Armement (DGA) via the Agence pour l'Innovation de D\'efense (AID). Sentinel-1 data was provided by the European Space Agency (ESA); water levels by the Canadian Hydrographic Service (CHS); and HFR data by Ocean Networks Canada (ONC). Many thanks go to Dr~Caro\-line~Paugam for her advices about storm surges.

\end{document}